\documentstyle[12pt,epsf]{article}
\begin{document}
\begin{center}
{\bf \Large Electromagnetic Production of Kaons on the Nucleon}
\end{center}
\begin{center}
T. MART\\
{\it Institut f\"ur Kernphysik, Johannes Gutenberg-Universit\"at, 
55099 Mainz, Germany}
\end{center}
\begin{center}
C. BENNHOLD, L. ALFIERI\\
{\it Center of Nuclear Studies, Department of Physics,\\
The George Washington University, Washington, D.C. 20052, USA}
\end{center}

\vspace{8mm}

\begin{center}
{\bf Abstract}
\end{center}
\begin{small}
Kaon photo- and electroproduction in all six isospin channels are
investigated by means of an isobaric model. It is found that  
existing models can only partially describe the current experimental data. 
Moreover, the extracted coupling constants are mostly inconsistent with 
the prediction of SU(3) symmetry or the information from hadronic 
reactions. In order to further analyze this problem, we introduce a 
form factor at the hadronic vertices $KNY$ (with $Y=\Lambda$ or 
$\Sigma$) and study its behavior qualitatively. Our result suggests the 
need of more accurate experiments, especially in the charged $\Sigma$
and the neutral $K^0$ channels.
\end{small}

\vspace{8mm}

The electromagnetic production of kaons has the potential to 
become a powerful tool
in the investigation of reactions involving strangeness, a  
degree of freedom not available with pions and eta mesons. 
The main problem faced by phenomenological models over the last 
thirty years, apart from the limited energy region,
is that there is little guidance as to which resonances to include in
the model. Furthermore,
 the extracted Born coupling constants are inconsistent with 
the SU(3) prediction or those yielded by the kaon-nucleon or 
hyperon-nucleon scattering, and the $\chi^2$ remains fairly large. 

In previous studies \cite{terry1,terry2} we have investigated kaon
photo- and electroproduction for the four $K\Sigma$ channels and found 
that the inclusion of the few available data for the reactions $\gamma p 
\rightarrow K^0 \Sigma^+$ and $\gamma n \rightarrow K^+ \Sigma^-$ 
in the fit can drastically reduce the leading coupling constants 
$g_{K\Lambda N}$ and $g_{K\Sigma N}$ in the process. This 
result suggests the need to include a form factor at the hadronic 
vertices, since such a form factor is expected to reduce the divergent 
Born terms at higher energies. However, the presence of a form factor 
at the $KNY$ vertices leads to the violation of gauge invariance, since 
different diagrams give different contributions of the form factor. 
There are several ways to remedy the situation, the simplest is to 
just multiply the whole amplitude with an overall form factor.
Another possibility would involve minimal 
substitution which was motivated by Ohta \cite{ohta}. 

In the present study we use the isobaric model developed in 
Ref.~\cite{terry1} with 
a slight modification in order to describe the elementary processes 
of photo- and electroproduction in all six isospin channels. 
To relate all production channels, we employ the isospin formalism 
and use some information on resonance decay widths \cite{terry1}. Our model 
consists of the standard Born terms along with the intermediate 
$K^*$-exchange.  Furthermore, we have incorporated the $N^*$ 
resonances $S_{11}$(1650) and $P_{11}$(1710) for $K\Sigma$ and 
$K\Lambda$ productions, as well as the $\Delta$ resonances $S_{31}$(1900) 
and $P_{31}$(1910) for the $K\Sigma$ channels. Our choice of resonances 
was guided by our goal to draw qualitative conclusions about the 
behavior of coupling constants with a simple model that contains 
as few parameters as needed to achieve a reasonable $\chi^2$.

\begin{table}[htb]
\begin{center}
\caption{The Born coupling constants from SU(3) prediction, fit to $K^+
         \Sigma^0$ data only (I) \protect\cite{benn}, fit to all 
         $K\Sigma$ data (II)
         \protect\cite{terry1}, fit to all $K\Sigma$ and $K\Lambda$ data
         (III), and fit to all data including the preliminary data
         \protect\cite{terry2} and using the hadronic form factor (IV).}
\begin{tabular}{lrrrrr}
\hline\hline\\
 &~~~~~SU(3)&~~~~~~~~~~~~~~I~~&~~~~~~~~~~~~II~~&~~~~~~~~~~~~~~III~~&
~~~~~~~~~~~~~~IV~~\\[1.5ex]
\hline\\
~~$g_{K\Sigma N}/\sqrt{4\pi}$& 1.09&2.72&0.13&0.53&1.13~~\\
~~$g_{K\Lambda N}/\sqrt{4\pi}$&$-3.74$&$-1.84$&0.51&$-2.10$&$-3.09$~~\\
~~$\Lambda_{\rm c}$ (GeV)&-&-&-&-&0.859~~\\
~~$N$&-&86&190&671&754~~\\
~~$\chi^{2}/N$&-&3.15&5.30&7.18&5.94~~\\
[1.5ex]
\hline\hline 
\end{tabular}
\label{th:terry:cc}
\end{center}
\end{table}

\begin{figure}[ht]
\centerline{\epsfxsize=10.5cm \epsffile{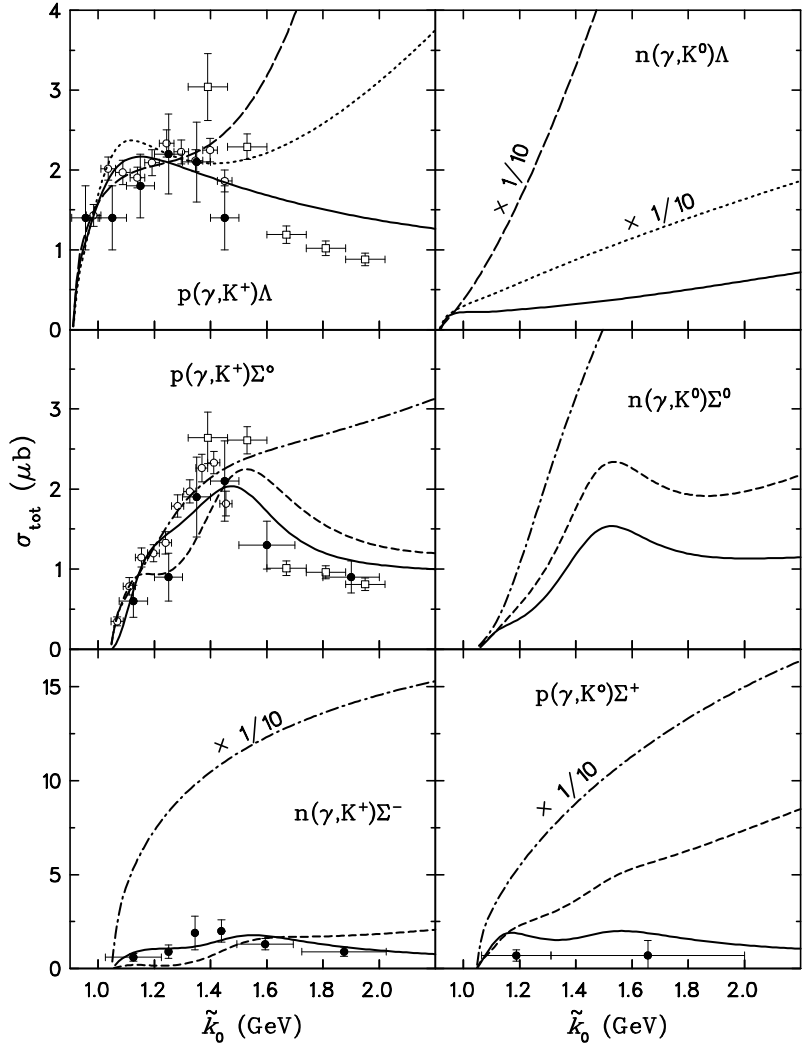}}
\caption{Total cross section in kaon photoproduction. The dotted (dashed)
         curve in $K\Lambda$ represents the model from 
         Ref.~\protect\cite{cota} (\protect\cite{saghai}). The dash-dotted
         (dashed) curve in $K\Sigma$ comes from set I (set II),
         while the solid curve fits all data in
         the $K\Lambda$ and $K\Sigma$ channels (set IV).}
\label{theory:tm:a}
\end{figure}

To obtain a qualitative understanding of the amplitude with a hadronic form
factor, we multiply the whole amplitude with a monopole form factor   
\mbox{$F(\Lambda_{\rm c},t) = (\Lambda_{\rm c}^2 - m_{K}^2)/ 
(\Lambda^2_{\rm c} - t)$}, where $\Lambda_{\rm c}$ represent the corresponding
cut off parameter. The result of our analyses along with that of former 
studies are shown in Table~\ref{th:terry:cc}. As shown in set IV of this
Table, the
 impact of the hadronic form factor is to increase the leading 
Born coupling constants to values consistent with the SU(3) prediction 
and to reduce the $\chi^2/N$ significantly using the 
entire data set. In Fig.~\ref{theory:tm:a}, we show the result for total 
cross sections of photoproduction in all six isospin channels. 
Obviously, the hadronic form factor can successfully suppress 
the divergence of the cross section at higher energies, whereas 
the $K^+\Lambda$ channel models from Refs.~\cite{saghai,cota} 
overestimate the cross section already at photon energy about 1.5 GeV. 
Besides diverging at higher energies in the $K^+\Sigma^0$ channel, 
the model of Ref.~\cite{benn} overpredicts experimental data in charged 
$\Sigma$ channels by up to two orders of magnitude.
Aside from the $n(\gamma,K^0)\Sigma^0$ channel, where no experimental 
data are available, Fig.~\ref{theory:tm:a} shows that a significant 
improvement to our former result \cite{terry1} has been made 
by our present model.
New and more accurate data, especially in the charged $\Sigma$ and 
the $n(\gamma,K^0)\Lambda$, as well as the $n(\gamma,K^0)\Sigma^0$ 
channels are therefore needed to improve the model.
In conclusion, we have investigated kaon photo- and electroproduction 
in all six isospin channels simultaneously and shown that the use 
of hadronic form factor can significantly improve the isobaric model, 
especially at higher energies.

\vspace{5mm}

The work of TM is supported by the Deutscher Akademischer Austauschdienst 
and the Deutsche Forschungsgemeinschaft (SFB 201), while the work of CB 
and LA is supported by the US DOE grant no. DE-FG02-95-ER40907.

\newpage

\end{document}